\documentstyle[12pt,aaspp4]{article}
\received{}
\revised{}

\lefthead{Vasconcelos, M.J., Jatenco-Pereira, V. and Opher, R.}
\righthead{Alfvenic Heating of accretion disks}

\begin{document}

\title{ALFVENIC HEATING OF PROTOSTELLAR ACCRETION DISKS}

\author{M. J. Vasconcelos, V. Jatenco-Pereira and R. Opher}

\affil{Instituto Astron\^{o}mico e Geof\'{\i}sico, Universidade de
S\~{a}o Paulo \\
Av. Miguel St\'efano 4200, 04301-904 S\~{a}o Paulo, SP, BRAZIL \\
jaque@iagusp.usp.br, jatenco@iagusp.usp.br, opher@iagusp.usp.br}


\begin{abstract}

We investigate the effects of heating generated by damping of Alfv\'en
waves on protostellar accretion disks. Two mechanisms of damping are
investigated, nonlinear and turbulent, which were previously studied
in stellar winds (\cite{jatenco...89a}, 1989b). For the nominal values
studied, $f=\delta v / v_{A}=0.002$ and $F=\varpi / \Omega_{i}=0.1$, where
$\delta v$, $v_{A}$ and $\varpi$ are the amplitude, velocity and average
frequency of the Alfv\'en wave, respectively, and $\Omega_{i}$ is the
ion cyclotron frequency, we find that viscous heating is more important
than Alfv\'en heating for small radii. When the radius is greater than
$0.5$ AU, Alfvenic heating is more important than viscous heating.  Thus,
even for the relatively small value of $f=0.002$, Alfvenic heating can be
an important source of energy for ionizing protostellar disks, enabling
angular momentum transport to occur by the Balbus-Hawley instability.


\end{abstract}

\clearpage

\section{Introduction}

Accretion disks are a powerful energy reservoir and are expected to be
found around pre-main sequence objects and active galactic nuclei and in
binary systems. One of the most challenging problems that still remains
concerning these objects is the nature of the mechanism by which they
transport angular momentum. A very promising mechanism was proposed some
time ago by \cite{balbus...91} (see also \cite{balbus...98}), based on a
previous studied instability (\cite{veli59}, \cite{chandra60}). An ambient
magnetic field, together with a differentially rotating system, can give
rise to a powerful instability that is very effective in transportating
angular momentum in the disk (e.g., \cite{stone...96}). However, because
of its magnetic nature, the instability needs a minimum ionization
density, $(n_{e}/n_{n}) \gtrsim 10^{-13}$, in order to be effective
(Gammie 1996).  The presence of this minimum ionization density in
protostellar accretion disks is uncertain. \cite{blaes...94} made a
linear study of the Balbus-Hawley instability (BHI) in a partially
ionized medium. \cite{hawley...98} did 3D numerical simulations and
found that, in a disk that is not totally ionized, when $ 0.01 <
(\gamma \rho _{i}/\Omega) < 100$ (where $\gamma \rho _{i}$ is the
collision frequency and $\Omega$, the orbital frequency of the disk),
there is angular momentum transport by BHI.  Greater heating in the
disk, in addition to that due to viscosity, will increase its degree of
ionization and a larger fraction of the disk will be susceptible to the
BHI. Our aim in this article is to analyze an extra source of plasma
heating for the ionization of the neutral part of the disk, allowing
the BHI to be effective in a greater part of it.

A system that is magnetized and turbulent implies the generation of
MHD waves.  These MHD modes can be damped by several mechanisms,
including collisions between ionized and neutral particles,
resistivity, electron viscosity, electron thermal conductivity, ion
viscosity (\cite{kulsrud...69}) and mode coupling (\cite{melrose80},
\cite{holzer...83}, \cite{zweibel...83}). Damping of MHD modes can
heat the gas. The damping of Alfv\'en waves has been considered to be
a heating mechanism for the solar corona (e.g., \cite{narain...95})
and in the heating of filaments in cooling flows in galaxy clusters
(\cite{friaca...97}).  It has also been considered as a source of
acceleration and heating of the solar wind (\cite{jatenco...89a},
\cite{jatenco...94}), as well as a generating mechanism of winds in
late-type giant stars (\cite{holzer...83}, \cite{jatenco...89b}).

There may be other sources of energy which can heat the disk, in addition
to viscous heating. For example, \cite{dalessio...99} showed that the
irradiation from the central star, as well as ionization by energetic
particles may be important for heating the disk.  The irradiation
is most important for R $\gtrsim$ 1 AU. As discussed by the authors,
stellar irradiation heats the dust, while the viscous dissipation heats
the gas. If the gas and the dust are well mixed, the disk will have the
same temperature for both components. Alfv\'en waves will heat the gas,
as will viscous dissipation. In principle, we can have these mechanisms
(stellar irradiation, viscous dissipation and Alfvenic heating) all
working together.

In general, the structure of a protostellar accretion disk is thought to
be as shown schematically in Figure \ref{fig1}. Region 1 is the boundary
layer between the disk and the star.  This region can, in principle,
extend up to the stellar surface. A difference in the rotational velocity
of the two objects (disk and star) causes heating. The rotational velocity
of the disk ($\sim 260$ km ${\rm s}^{-1}$) is much greater than that of
the star ($\sim 10$ km ${\rm s}^{-1}$) and a large amount of energy is
dissipated in a thin disk layer (\cite{lynden...74}).  However, if the
star has a strong magnetic field ($\sim 1$ kG; \cite{wang96}), it can
disrupt the disk, with the accretion proceeding along the field lines
(\cite{ghosh...79a}, b; \cite{konigl91}). The magnetospheric accretion
model accounts for many observational properties of T Tauri stars, such
as the lack of a strong ultraviolet flux and the presence of blueshifted
permitted lines (\cite{edwards97}).  A controversial point in this class
of models is the precise location of the place where the magnetic field
truncates the disk (\cite{hart98}).  Depending upon where this point
(or region) is, the star can be spun up or down.  Several wind models
depend upon the determination of this location (e.g., \cite{shu...94},
2000). Region 2 of the disk extends from $\sim 0.01$ AU
to $\sim 1$ AU and is relatively close to the star. In this region,
for 0.01 AU $<$ R $<$ 0.1 AU,
the temperature is above $10^3$ K, which is sufficient to couple the
magnetic field to the gas, and it is likely that the BHI is present
here. However, when the temperature falls below $10^3$ K, the role of BHI
is uncertain. 
Region 3 in the figure extends from 1.5 AU to $\sim$ 5AU.
\cite{gammie96} claims that for R $\gtrsim$
0.1 AU the disk separates into three regions. Two regions, above and
below the midplane of the disk are ionized by cosmic rays ataining the
necessary temperature to couple the magnetic field with the gas. Cosmic
rays, however, do not penetrate deep enough to heat the midplane region
and the magnetic field is thought to decouple from the gas preventing
the BHI from operating. \cite{igea...99} believe that X-rays from the
central star are more effective in ionizing the disk than are cosmic rays.
In region 4, the density of the disk is small enough for the disk to be
totally ionized again. 
Somewhere in this region, the disk becomes optically thin and submillimeter
observations provide information about the disk mass.

\placefigure{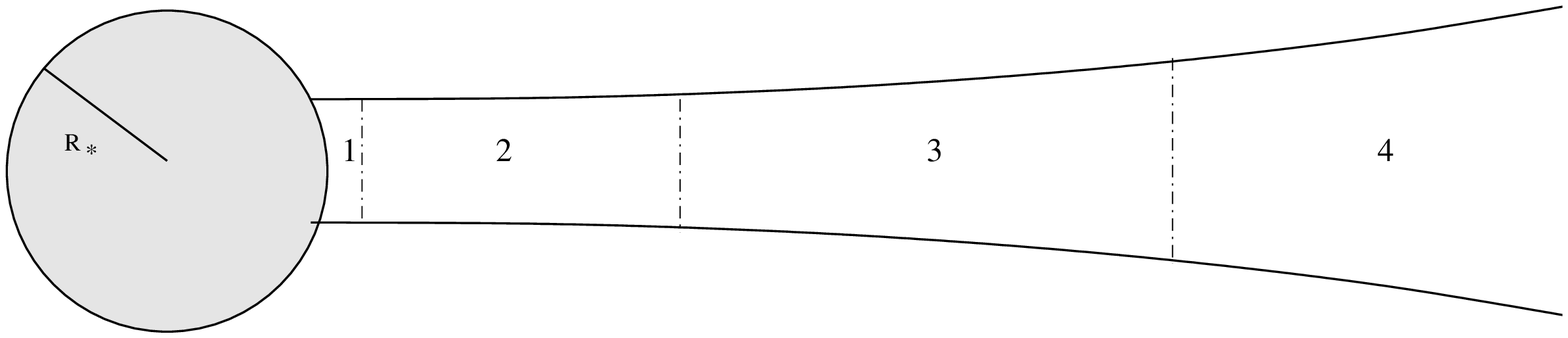}

In this article, the extra source of heating for ionizing the neutral
part of a protostellar accretion disk is assumed to be due to the damping
of Alfv\'en waves. The damping mechanisms studied are nonlinear damping
(section \ref{nonlinear}) and turbulent damping (section \ref{turbulent}).
For both damping mechanisms, we considered the cases of constant Alfvenic
heating, exponentially varying density and temperature, as well as a
layered model.  We have made our calculations for the region that extends
from 0.1 AU to $\sim$ 100 AU.
The results and discussions are presented in
section \ref{resedisc}, our conclusions are given in section \ref{conc}.

\section{Alfvenic Heating} 

In order to quantify the amount of Alfvenic heating in a protostellar
accretion disk we study the equations (in cylindrical coordinates) that
describe a geometrically thin, optically thick, non-magnetized accretion
disk (\cite{pri81}, \cite{hart98}):

\begin{equation} \label{fluxo}
\frac{dF_{z}}{dz}=\frac{9}{4}\nu \rho \frac{GM_{\ast}}{R^{3}},
\end{equation}

\begin{equation} \label{nisigma}
\nu \Sigma = \frac{\dot{M}}{3\pi}\left[1-\left(\frac{R_{\ast}}{R}\right)^{1/2}\right],
\end{equation}
where $F_{z}$ is the energy flux in the vertical z direction, $\nu$
is the viscosity, $\rho$ is the fluid density, $G$ is the gravitational
constant, $M_{\ast}$ is the mass of the central star, $R$ is the distance
from the central star, $\Sigma$ is the surface density and $R_{\ast}$,
the radius of the central star.

Integrating equation (\ref{fluxo}), we obtain

\begin{equation} \label{Fz}
F_{z}=\frac{9}{4}\nu \Sigma \frac{GM_{\ast}}{R^{3}},
\end{equation}
where we made use of the relation $\Sigma=\int \rho dz$.  Substituting
equation (\ref{nisigma}) into equation (\ref{Fz}), we obtain the total
flux of energy $(\rm{erg \; cm^{-2} \; s^{-1}})$ generated by viscous
dissipation in the disk as a function of the disk radius:

\begin{equation} \label{Dvis}
F_{z}=D_{Tot}=D_{vis}=\frac{3GM_{\ast}\dot{M}}{8\pi}\left[1-
\left(\frac{R_{\ast}}{R}\right)^{1/2}\right]R^{-3},
\end{equation}
where $\dot{M}$ is the accretion rate of the disk.  Equation (\ref{Dvis})
is the gravitational potential energy liberated by the accretion flux.
Since, in the present paper, we are examining the damping of Alfv\'en
waves as an extra source of energy for heating the disk, we have

\begin{equation} \label{Dtot}
D_{Tot}=D_{vis} + D_{A},
\end{equation}
where $D_{A}$ is the energy generated by the damping of Alfv\'en waves
in the disk.

It is known that the nonlinear growth of the Balbus-Hawley instability
results in turbulence (\cite{stone...96}) and so, we can have MHD modes
traveling throughout the disk. These MHD modes can be damped by several
mechanisms (e.g., \cite{holzer...83}, \cite{zweibel...83}). Here, we deal
with two of these mechanisms: nonlinear damping and turbulent damping.
These two damping mechanisms have been previously investigated in a
series of articles related to stellar winds by Jatenco-Pereira \& Opher
(1989a, 1989b) and Jatenco-Pereira et al.  (1994). Here we investigate
these damping mechanisms in relation to protostellar accretion disks.

High amplitude waves cause nonlinear mode couplings and, in this process,
the resulting modes are rapidly damped. In general, the waves that result
from the damping processes are sound waves; thus, the nonlinear Alfvenic
energy is, in general, converted into thermal energy.  If the Alfv\'en
waves are not in phase, their energy is transfered both to the fast and
slow magnetosonic modes (e.g., \cite{holzer...83}).  In a high $\beta$
plasma ($\beta = 8\pi P/B^{2} > 1$, where $P$ is the gas pressure and
$B^{2}/8\pi$, the magnetic field pressure), the mode coupling of the
Alfvenic mode to the slow magnetosonic mode can be very strong because the
phase velocities of the modes are essentially the same (\cite{melrose80}).

Nonlinear damping of Alfv\'en waves requires waves of great amplitude
$\delta B/B$ (where $\delta B$ is the amplitude of the magnetic field of
the wave and $B$ is the strength of the magnetic field of the medium). The
turbulent damping mechanism requires the presence of turbulent cells or
vortices. If we have a turbulent medium in which $\delta B/B$ is large, in
principle, we can have both mechanisms working together. In this paper
we treat each mechanism separately.

In the next sub-section, we consider the nonlinear damping mechanism of
Alfv\'en waves.  (The damping rates used are obtained from Gon\c calves
et al. (1998), hereafter GJO).

\subsection{Nonlinear damping of Alfv\'en waves} \label{nonlinear}

From equation (\ref{Dtot}), the total amount of heating in a magnetized
accretion disk, assumed to be due only to Alfv\'en damping and viscosity,
is

\begin{equation} \label{Dtot2}
D_{Tot}=D_{vis} + \int_{-h}^{+h} H_{A}dz=D_{vis} + 2\int_{0}^{+h} H_{A}dz,
\end{equation}
where $h$ is the half thickness of the disk, $H_{A}=\Phi _{w}/L_{A}$
is the rate of Alfvenic heating, $\Phi _{w}$ is the Alfv\'en wave flux,
$L_{A}=v_{A}/\Gamma_{w}$ is the damping length, $v_{A}=B/\sqrt{4\pi
\rho}$ is the Alfv\'en velocity and $\Gamma_{w}$, the damping rate of
the waves. The nonlinear damping rate is given by

\begin{equation} \label{GammaNL}
\Gamma _{NL}=\frac{1}{4}\sqrt{\frac{\pi}{2}}\xi\varpi\left(\frac{c_{s}}
{v_{A}}\right)\frac{\rho \langle \delta v^{2}\rangle }{B^{2}/8\pi},
\end{equation}
where $\xi$ is a constant with a value between $5$ and $10$, $\varpi$ is
the characteristic frequency of the waves, $c_{s}$ is the sound velocity
and $\rho \langle \delta v^{2}\rangle /(B^{2}/8\pi)$, the energy density
of the waves divided by the magnetic energy density (\cite{dede...98}).
The energy density of the waves can also be written as (\cite{whang97})

\begin{equation} \label{whangen}
\epsilon_{Alfv\acute{e}n}=\frac{1}{2}\rho \langle \delta v^{2}\rangle +
\frac{1}{8\pi}\langle \delta B^{2}\rangle.
\end{equation}

The two terms on the right side of the equation (\ref{whangen}) are
generally equal, so that

\begin{equation} \label{EWhang}
\epsilon_{Alfv\acute{e}n}=\rho\langle \delta v^{2}\rangle .
\end{equation}

Using the equations above, the wave flux can be expressed as

\begin{equation} \label{fluxoalf}
\Phi _{w}=\rho\langle \delta v^{2}\rangle v_{A},
\end{equation}
where $\langle \delta v^{2} \rangle$ is proportional to the degree of
turbulence of the system.

>From equations (\ref{GammaNL}) and (\ref{fluxoalf}), we obtain the nonlinear
Alfvenic heating rate:

\begin{equation} \label{HNL1}
H_{NL} = \frac{\rho \langle \delta v^{2} \rangle v_{A}}{v_{A}} \frac{1}{4}
\sqrt{\frac{\pi}{2}}\xi\varpi\left(\frac{c_{s}}
{v_{A}}\right)\frac{\rho \langle \delta v^{2}\rangle }{B^{2}/8\pi}.
\end{equation}

The adiabatic sound velocity is $c_{s}=\sqrt{\gamma P/\rho}$ and, for an
ideal gas, $P=\rho \Re T/\mu$, where $\Re$ is the universal gas constant.
We define the parameter $f$ by

\begin{equation} \label{deff}
\langle \delta v^{2} \rangle = f^{2}v_{A}^{2}.
\end{equation}

The average frequency $\varpi$ can be written as

\begin{equation} \label{defwf}
\varpi=F\Omega_{i}=F\frac{eB}{m_{i}c}, 
\end{equation}
where $F (<1)$ is a free parameter, $\Omega_{i}$ is the ion-cyclotron
frequency and $e$, $m_{i}$ and $c$ are the electron charge, the ion mass
and the velocity of light, respectively.

Substituting equations (\ref{deff}) and (\ref{defwf}) in equation (\ref{HNL1}),
we have

\begin{equation} \label{HNL2}
H_{NL}=\frac{\sqrt{2}}{8}f^{4}F\frac{e\xi}{m_{i}c}\left(\frac{\gamma \Re}{\mu}
\right)^{1/2}(\rho T)^{1/2}B^{2}.
\end{equation}

We then have, from equations (\ref{Dtot2}) and (\ref{HNL2})

\begin{equation} \label{DNL}
D_{NL}=2 \int_{0}^{h} H_{NL}dz = \frac{\sqrt{2}}{4}f^{4}F\frac{e\xi}
{m_{i}c}\left(\frac{\gamma \Re}{\mu}\right)^{1/2}\int_{0}^{h}
(\rho T)^{1/2}B^{2}dz.
\end{equation}

\cite{gammie96} proposed a model in which a protostellar accretion disk
has different layers with different ionization rates. Near the star
$(R \lesssim 0.1 {\rm AU})$, the temperature is greater than $10^{3}$
K and, thus, collisional ionization is sufficient to couple the gas to
the magnetic field. Beyond $R=0.1 {\rm AU}$, the temperature drops and
only small layers above and below the central plane of the disk (called
``active layers'') are able to sustain the Balbus-Hawley instability
by cosmic ray ionization. The central layer, called the ``dead zone'',
is neutral, with insufficient ionization to enable coupling.  The active
layers have a constant surface density, which depends on the cosmic-ray
ionization rate, a poorly known quantity. \cite{gammie96} adopted the
interstellar value (\cite{spitzer...68}) for this rate and obtained
$\Sigma_{a}=10^{2} \; {\rm g} \; {\rm cm}^{-2}$, for the surface density
of the active layers. Thus, if we assume that beyond the radius of $0.1$
AU the surface density of the active layers is constant, we obtain from
equation (\ref{DNL})

\begin{equation} \label{DNL_bey}
D_{NL}=\frac{\sqrt{2}}{4}f^{4}F\frac{e\xi}
{m_{i}c}\left(\frac{\gamma \Re}{\mu}\right)^{1/2}\Sigma^{1/2}_{a} \int_{0}^{h}
T^{1/2}B^{2}z^{-1/2}dz, \; (R > 0.1 AU),
\end{equation}
where the relation $\Sigma=\rho z$ was used and it is assumed that the
thickness of the central neutral region is very small compared to the
thickness of the active layers (Gammie 1996).

In quantifying the Alfvenic heating of a protostellar accretion disk,
we consider various cases.  The magnetic field in all of these cases,
is held constant in the vertical direction, as was done in various
previous papers (e.g., \cite{stone...94}; \cite{armi98}; \cite{wardle99}).
First, we assume that the density and the temperature are constant
throughout the disk (section \ref{CNAH}).  Then, we vary the density but
still assume an isothermal disk (section \ref{DVAH}). In the next case, we
vary the temperature as well as the density (section \ref{TVAH}).  Finally
(section \ref{layNL}), we consider the layered model of \cite{gammie96}.
In the following subsections, we assume an average effective value for
$f$ and $F$ in the disk.

\subsubsection{Constant Nonlinear Alfvenic Heating} \label{CNAH}

Considering the density, temperature and magnetic field to be constant
throughout the disk, we integrate equation (\ref{DNL}) and obtain

\begin{equation} \label{DNL3}
D_{NLconst}=\frac{\sqrt{2}}{4}f^{4}F\frac{e\xi}{m_{i}c}
\left(\frac{\Re}{\mu}\right)^{1/2}T_{d}^{1/2}B^{2}\Sigma^{1/2}h^{1/2},
\end{equation}
where we used $\gamma=1$, since the temperature is constant everywhere.
The temperature $T_{d}$, adopted above, is obtained by assuming that the
disk radiates as a blackbody and that the only source of dissipation
is viscous heating. From $D_{vis}=\sigma T^{4}_{d}$ (where $\sigma$
is the Stefan-Boltzmann constant), we obtain

\begin{equation}
T_{d}^{4}=\frac{3GM_{\ast}\dot{M}}{8\pi\sigma}
\left[1-\left(\frac{R_{\ast}}{R}\right)^{1/2}\right]R^{-3}.
\end{equation}

%
%




\subsubsection{Nonlinear Alfvenic Heating with Exponentially Varying density} \label{DVAH}

According to the model of a stationary, geometrically thin accretion
disk (\cite{pri81}), if the temperature in the vertical direction
is held constant, the resultant density profile is proportional to
$exp(-z^2)$. However, in a turbulent disk, it is expected that the
decrease in density in the vertical direction is less dramatic, so that
we assumed that the density falls as $exp(-z)$.  Thus, considering a
disk with an exponentially varying density, $\rho = \rho_{c}e^{-z/h}$,
equation (\ref{DNL}) is given by

\begin{equation} \label{DNLrho}
D_{NL\rho var} \simeq \frac{\sqrt{2}}{5}f^{4}F\frac{e\xi}{m_{i}c}
\left(\frac{\gamma \Re}{\mu}\right)^{1/2}(\rho_{c}T_{d})^{1/2}B^{2}h, 
\end{equation}
where we again used $\gamma=1$. 

\subsubsection{Nonlinear Alfvenic Heating with Exponentially Varying
density and Temperature} \label{TVAH}

In order to obtain a consistent temperature profile in the vertical
direction, it is necessary to solve the equations that describe the disk
in the z direction (e.g., \cite{dalessio96}; \cite{dalessio...98}, 1999).
This calculation requires a detailed treatment of the transport of the
radiation and the advection of material by the BHI, which is beyond the
scope of this paper. For simplicity, we have assumed that the temperature
of the disk in the vertical direction falls off exponentially.  So,
in the case of a disk with both a density and temperature variation
$(T=T_{c}e^{-z/h})$ the Alfvenic heating is given by

\begin{equation} \label{DNLvar}
D_{NLvar} \simeq \frac{3\sqrt{2}}{20}f^{4}F\frac{e\xi}{m_{i}c}
\left(\frac{\gamma \Re}{\mu}\right)^{1/2}(\rho_{c}T_{c})^{1/2}B^{2}h.
\end{equation}

%
%
\subsubsection{The Layered Model} \label{layNL}

In this case, to quantify the amount of Alfvenic heating of the disk,
we adopt the model of \cite{gammie96}.
%
%
We consider only the region beyond $0.1$ AU, where it is generally assumed
that the disk is divided into three regions with different ionization
densities. Adopting $T=T_{c}$, together with a constant magnetic field,
and integrating equation (\ref{DNL_bey}), we obtain

\begin{equation} \label{DNL_bey_2}
D_{NLlay}=\frac{\sqrt{2}}{2}f^{4}F\frac{e\xi}
{m_{i}c}\left(\frac{\gamma \Re}{\mu}\right)^{1/2}(\Sigma_{a}T_{c}\; h)^{1/2}
B^{2}
\end{equation}     


%
%
%
%

\subsection{Turbulent Damping} \label{turbulent}

A damping length, $L_T$, can describe the absorption of turbulent
Alfv\'en waves over a broad range of Alfv\'en periods. Similar to a
Kolmogorov spectrum, a turbulent cascade transfers wave energy from large
to small scale regions. The dissipation of energy through turbulence is
governed by the small scale linear absorption mechanism.  Noting the
similarity of $P_{B} \propto k^{-5/3}$ and Kolmogorov turbulence in
ordinary fluids\footnote{Where $P_{B}$ is the observed magnetic power
spectra in the solar wind and $k$ is the wavenumber (\cite{hollweg87}).},
 the volumetric heating rate associated with the cascade is $H_{T} =
\rho {\langle \delta v^2 \rangle}^{3/2}/L_{corr}$, where $L_{corr}$ is
a measure of the transverse correlation length (\cite{hollweg86}). In
terms of the damping length,

\begin{equation}
L_T = L_{corr} \, v \, {\langle \delta v^2 \rangle}^{-1/2} \;,
\end{equation}
where $L_{corr} \propto B^{-1/2}$ (\cite{hollweg86}) and $v$ is the
fluid velocity.

In the present work, we treat $L_T$ as an independent damping process,
although it may be connected with non-linear damping.

Then, the Alfvenic heating rate is given by (\cite{dede...98})

\begin{equation}
H_{T}=\frac{\Phi_{w}B^{1/2}\langle \delta v^{2}\rangle^{1/2}}{v_{A}} = 
\rho B^{1/2}\langle \delta v^{2}\rangle^{3/2}.
\end{equation}

Using equation (\ref{deff}), we obtain

\begin{equation} \label{Ht}
H_{T}=\rho B^{1/2}f^{3}v^{3}_{A}=\left(\frac{f}
{\sqrt{4\pi}}\right)^{3}B^{7/2}\rho^{-1/2}.
\end{equation}

In this case, the damping heating rate does not depend on temperature.
Similar to our analysis in section \ref{nonlinear}, we consider three
cases in order to obtain the total dissipation due to turbulent damping of
Alfv\'en waves:  1) a disk with a constant density (section \ref{CTH});
2) a disk with a variable density (section \ref{DVTH}) and 3) a layered
disk (section \ref{layTH}).

\subsubsection{The Constant Turbulent Damping} \label{CTH}

We now consider a disk in which the density and the magnetic field are
constant throughout.  Integrating equation (\ref{Ht}), we obtain

\begin{equation} \label{Dtconst}
D_{Tconst}=2\left(\frac{f}{\sqrt{4\pi}}\right)^{3}
\left(\frac{B^{7}h^{3}}{\Sigma}\right)^{1/2}.
\end{equation}

%
%
\subsubsection{Variable Density} \label{DVTH}

Assuming that the density is varying exponentially with height
$(\rho=\rho_{c}e^{-z/h})$, we can integrate equation (\ref{Ht}), obtaining

\begin{equation} \label{Dtvar}
D_{Tvar}=2.6\left(\frac{f}{\sqrt{4\pi}}\right)^{3}\left(\frac{B^{7}}{\rho_{c}}
\right)^{1/2}h.
\end{equation}

\subsubsection{Layered Disk} \label{layTH}

Integrating equation (\ref{Ht}), using the layered model of constant surface
density $(\rho=\Sigma_{a}/z)$ (section \ref{layNL}), we obtain the total
amount of heating for the turbulent damping of Alfv\'en waves,

\begin{equation} \label{Dtlay}
D_{Tlay}=\frac{4}{3}\left(\frac{f}{\sqrt{4\pi}}\right)^{3}
\left(\frac{B^{7}h^{3}}{\Sigma_{a}}\right)^{1/2}.
\end{equation}

This equation is valid for the region beyond $0.1$ AU, where it is
assumed that there are active layers, created by cosmic ray ionization.

\section{Results and Discussions} \label{resedisc}

In Table \ref{tab1}, we show the values of the parameters of the disk
that are used to calculate the different types of Alfvenic heating
for region 2, except for those related
to the layered model of \cite{gammie96} $(D_{NLlay}$ and $D_{Tlay})$.
These values (except the magnetic field) are obtained using a model of a
stationary, geometrically thin, optically thick, $\alpha$ accretion disk
(e.g., \cite{pri81}, \cite{frank...92}, \cite{shak...73}) with the opacity
law of \cite{bell...94} in the iron-silicate regime\footnote{In this 
regime, the opacity is dominated by iron and silicate grains. The 
corresponding range of temperature is between $\sim 10^{3}$ K to 
$203$ K (Gammie 1996).}.
The magnetic field is obtained from the model of \cite{gammie96}.

\placetable{tab1}

In Table \ref{tab2}, we show the disk parameters obtained from the
model presented by \cite{gammie96} for region 2. With the parameters 
shown in Table
\ref{tab2}, we calculate the layered Alfvenic heating for region 2. 
The values
of both models are obtained considering a disk surrounding a T Tauri
star of $0.5 M_{\sun}$, which is accreting mass at a rate of $10^{-8}$
${\rm M}_{\sun}$ ${\rm year}^{-1}$. We adopt a value of $0.6$ for the
mean molecular weight $\mu$.

\placetable{tab2}

Temperatures associated with each dissipated energy mechanism may be
calculated, assuming that the disk radiates like a blackbody. Thus, there
is the following temperature associated with $D_{NLconst}$, for example:

\begin{equation} \label{Tcalc}
T_{const}=\left(\frac{D_{NLconst}}{\sigma}\right)^{1/4},
\end{equation}
where $T_{const}$ is the temperature and $\sigma$, the Stefan-Boltzmann
constant.

In Figure 2, we plot the temperatures associated with nonlinear
Alfvenic dissipation for disk region 2, calculated from equation
(\ref{Tcalc}) with $f=0.002$ (eq. [\ref{deff}]), $F=0.1$
(eq. [\ref{defwf}]) and $\xi$=5. Figure 2a shows the temperatures
associated with each kind of dissipation. $T_{vis}$ is obtained from
the viscous dissipation of the stationary, geometrically thin accretion
disk model (Table \ref{tab1}). For comparison, we plot $T_{irr}$,
the temperature associated with the model of the irradiated disk
(\cite{dalessio...98}). Plotted values in the figure were obtained from
Figure 3 of D'Alessio et al. (1998). The temperature calculated using
Alfvenic dissipation in the layered model ($T_{lay}$, shown as circles
in Fig. 2a) is highest for $R \geq 0.2$ AU. The other temperatures
($T_{const}$, $T_{var}$, $T_{\rho var}$) do not differ from one another
very much. $T_{vis}=(D_{vis}/\sigma)^{1/4}$ decreases with increasing
radius, becoming less than the other temperatures for $R > 0.4$
AU. The temperature obtained from the irradiated model is greater than
$T_{const}$, $T_{\rho var}$, $T_{var}$ for all radii considered.  We plot
the total temperatures obtained from our models together with $T_{vis}$
in Figure 2b. They are calculated from the following relations:

\begin{eqnarray} 
T_{tot1}=[T^{4}_{d} + T^{4}_{const}]^{1/4}=\left[\frac{D_{vis}}{\sigma} + 
\frac{D_{NLconst}}{\sigma}\right]^{1/4}, \\
T_{tot2}=[T^{4}_{d} + T^{4}_{\rho var}]^{1/4}=\left[\frac{D_{vis}}{\sigma} + 
\frac{D_{NL\rho var}}{\sigma}\right]^{1/4}, \\
T_{tot3}=[T^{4}_{d} + T^{4}_{var}]^{1/4}=\left[\frac{D_{vis}}{\sigma} + 
\frac{D_{NLvar}}{\sigma}\right]^{1/4}, \\
T_{tot4}=[T^{4}_{d} + T^{4}_{lay}]^{1/4}=\left[\frac{D_{vis}}{\sigma} + 
\frac{D_{NLlay}}{\sigma}\right]^{1/4}. 
\end{eqnarray}

We note the increase in temperature. Highest temperatures are again
obtained when the layered model is used. The different density and
temperature profiles, used to obtain the other Alfvenic dissipations
$(D_{NLconst}$, $D_{NL\rho var}$ and $D_{NLvar}$), do not affect the final
total temperature obtained. They are practically equal for all radii.

In Figure 2c, we plot the total temperatures, taking into account the
temperature from the irradiated model. The increase in temperature
is much more significant than when only viscous dissipation is
considered. Again, the density and temperature profiles do not alter
the final temperature apreciably. The highest temperature is obtained
from the layered model. $T_{irr}$ dominates the total temperature values,
$T_{T1}$, $T_{T2}$, $T_{T3}$, which indicates that irradiation may be more
important than the Alfvenic heating mechanisms which produced $T_{const}$,
$T_{\rho var}$, $T_{var}$ for this part of the disk.

\placefigure{fig2}

We further extend our calculations to include a larger amount
of the protostellar disk. We divide the disk according to the different
opacities laws, as given by \cite{gammie96}. We make our calculations
for two additional regions (regions 3 and 4 of Figure 1).
region 3 extending from 1.5 AU to 5.0 AU and
region 4 extending from $\sim 5$ AU to $\sim$ 100 AU. The opacity law
for region 3 is given by $\kappa = 2 \times 10^{16} {\rm T}^{-7}
{\rm cm}^{2}{\rm g}^{-1}$ and for region 4, by $\kappa = 2 \times 10^{-4}
{\rm T}^{2}{\rm cm}^{2}{\rm g}^{-1}$ (Bell \& Lin 1994). 

In Figure 3a, we show the temperatures obtained from nonlinear heating 
models, calculated for region 3 (1.5 AU $\leq$ R $\leq$ 5 AU). For this
region, $T_{vis}$ (showed again as up triangles in figure) is the
smallest temperature. The temperature obtained for the layered
model (circles) is greater than that obtained for other models 
for all radii considered.
The photospheric temperature obtained from the irradiated model
is compared with $T_{var}$ (stars and squares in the figure, respectively).
In this figure, we note a small separation between the different
profiles of temperature and density.
 
In Figure 3b, we plot the temperatures obtained from nonlinear
heating models for region 4, which extends from $\sim$ 5 AU to
$\sim$ 100 AU. We note small changes compared with Figure 3a. We can observe
an inversion of the order of the temperatures. For this case, $T_{var}$
is less than $T_{const}$ and $T_{\rho var}$. However, this inversion
is not very dramatic. $T_{lay}$, again, is the highest temperature
and $T_{irr}$ does not differ very much from the Alfvenic temperatures
$T_{const}$ and $T_{\rho var}$.
In this part of the disk, the presence of layers with different degrees of
ionization is less certain. Due to the small values of density,
the penetration of cosmic rays can be deeper, perhaps, sufficiently
to ionize this entire disk region.

\placefigure{fig3}

%
%
%
In Figure 4a, we show the turbulent Alfvenic heating and viscous
dissipation.  For all radii, Alfvenic heating is greater than the viscous
heating.  This means that the turbulent mechanism for damping can be much
more efficient to dissipate energy than the viscous mechanism. Moreover,
we can see that, in this case, the turbulent Alfvenic heating calculated,
using the layered model $D_{Tlay}$ (eq. [\ref{Dtlay}]), is less
than both the Alfvenic heating with constant density $D_{Tconst}$
(eq. [\ref{Dtconst}]) and the Alfvenic heating with exponentially
variable density $D_{Tvar}$ (eq. [\ref{Dtvar}]). However, $D_{Tconst}$
and $D_{Tvar}$ are practically the same for all radii. This means that
turbulent damping is not very dependent on density.
In Figure 4b, we plot the turbulent Alfvenic heating and viscous 
dissipation for region 3 (1.5 AU $<$ R $<$ 5.0 AU). We note that
viscous dissipation is very small, when compared with turbulent
dissipation. For $D_{\rho var}$ and $D_{const}$, there is a tendency
for an increase at the beginning but after R $\sim$ 3 AU this tendency
is reverted and the energy decreases. Again, $D_{lay}$ is
not very significant, when compared with the two other turbulent
dissipations. In Figure 4c, we show the turbulent dissipated energies
for region 4 (5 AU $\lesssim$ R $\lesssim$ 100 AU).Again, turbulent 
dissipation is greater than viscous dissipation.


\placefigure{fig4}

We know that an important parameter for disk models is the electron
fraction, $x=n_e/n_n$, where $n_e$ is the number density of electrons
and $n_n$, the number density of the neutral specie.  \cite{gammie96}
calculated an upper limit for this fraction, assuming ionization by
cosmic rays. He obtained
 
\begin{equation} \label{xgammie}
x=\left(\frac{\zeta}{\beta n_H}\right)^{1/2}=1.6 \times 10^{-12}
\left(\frac{T}{500 K}\right)^{1/4} \left(\frac{\zeta}{10^{-17} \rm{s}^{-1}}
\right)^{1/2} \left(\frac{n_H}{10^{13} \rm{cm}^{-3}}\right)^{-1/2},
\end{equation}
where $\zeta$ is the cosmic-ray ionization rate, $\beta$ is the
recombination coefficient and $n_H$, the number density of hydrogen.
This expression is valid for $T < 10^3$ K.  Using equation (\ref{xgammie})
and $T_{d}$, $\Sigma$ and $h$ from Table \ref{tab2}, we find that the
electron fraction increases with radius. However, in order to evaluate
$x$, \cite{stone...00} used an expression derived from the Saha equation,
which took into account the thermal ionization of potassium:
 
\begin{equation}
x=a^{1/2}T^{3/4}\left(\frac{2.4 \times 10^{15}}{n_n}\right)^{1/2}
{\rm exp}(-50,370/2T),
\end{equation}
where $a$ is the abundance of potassium. In this case, $x$ decreases
with radius (The expression is valid for $T < 2000 K$). It is possible
that, for the range of radii and temperature considered in this paper,
the expression used by \cite{gammie96} for $x$ is the most likely to be
valid. However, the determination of $x$ is still very uncertain.

\section{Conclusion} \label{conc}

In this paper, we analysed the important role that the damping of
Alfv\'en waves may possibly play in the inner and intermediate regions of
protostellar accretion disks $(0.1 \leq R (\rm{AU}) \lesssim 100)$. 
We considered two damping mechanisms: nonlinear and turbulent.

We considered various profiles of density and temperature, with a
constant magnetic field, in quantifying the heating generated by the
damping of the waves. Concerning nonlinear damping, in the 
first case, the density and temperature
are held constant throughout the disk.  The resultant heating is called
$D_{NLconst}$. In the second case, the density is exponentially varying,
while the disk is isothermal ($D_{NL\rho var}$).  Both temperature and
density are exponentially varying in the third case, $D_{NLvar}$. The
heating $D_{NLlay}$, the fourth case, is obtained when nonlinear damping
of Alfv\'en waves occurs in the disk model of \cite{gammie96} (the layered
model). We note that $D_{NLlay}$  is greater than the viscous dissipation,
$D_{vis}$ for $R \geq 0.2 \rm{AU}$. We see that, for R $>$ 0.5 AU
all the forms of nonlinear Alfvenic heating are greater than
viscous heating.  When the damping occurs by turbulent mechanisms, all
the resultant turbulent Alfvenic dissipations are greater than viscous
dissipation for all considered radii.  Figure \ref{fig2} shows the
calculated temperatures for our model for region 1 (0.1 AU $\leq$ R
$\leq$ 1.4 AU). We note an increase in temperature for this region
in Figure 2b.  Thus, we conclude that, when viscous dissipation
is insufficient to ensure the necessary degree of ionization for BHI
to occur, the damping of Alfv\'en waves can be an alternative source of
energy for ionization.
When taking into account the irradiation from the central
star, the increase in temperature is much greater than when only the
viscous dissipation is considered. Alfvenic heating in the layered model
is very significant, greater than all other dissipations considered.

In Figures 3a and 3b we show the temperatures obtained from the
nonlinear heatings for the outer parts of the disk. Again, the same
order for the temperatures remains, with little differences shown.
The viscous dissipation becomes smaller and less significant for
increasing radii. The irradiation is a rather important source
of heating, comparable with nonlinear Alfvenic $D_{const}$,
$D_{\rho var}$ and $D_{var}$. In Figures 4b and 4c we see that
turbulent dissipation is very efficient for heating the disk.
We thus conclude that Alfvenic heating can be important in the 
outer parts of the disk.  Considering only viscous
dissipation, the temperatures reached are not sufficient to ensure the
development of BHI and, consequently, of turbulence. If we take into
account the irradiation from the central star, temperatures
can reach values which ensure the necessary coupling between the gas 
and the magnetic field.

In the present paper, it is to be noted that we used $f(\equiv \delta
v/v_{A})= 0.002$ and $F(\equiv \varpi/\Omega_{i})=0.1$. Nonlinear Alfv\'en
heating is proportional to $f^{4}F$ and turbulent Alfv\'en heating,
to $f^{3}$. Increasing (decreasing) $f$ and $F$ from these nominal
values increases (decreases) the Alfv\'en heating.  We have shown,
in this paper, that even for the relatively small value of $f=0.002$,
Alfvenic heating can be important in protostellar disks.

\acknowledgements
M.J.V. would like to thank the Brazilian agency FAPESP for financial
support. V.J.P and R.O.  thank the Brazilian agency CNPq for partial
support. The authors would like to also thank the project PRONEX/FINEP
(No. 41.96.0908.00) for partial support.

\newpage
\figcaption[f1.eps]{Sketch of a protostellar accretion disk.
Region 1 corresponds to the boundary between the disk and the star. Region
2 extends from $\sim 0.01$ AU to $\sim 1$ AU. Region 3 extends
from $\sim 1$ AU to $\sim 5$ AU. Beyond this region, region 4,
the disk becomes optically thin. We make our calculations
for regions 2,3, and 4, extending from 0.1 AU to $\sim$ 100 AU. 
(See text for a more detailed discussion.)
\label{fig1}}

\figcaption[f2a,b,c.eps]{Temperatures associated with nonlinear Alfvenic
dissipation, calculated for 0.1 AU $<$ R $<$ 1.4 AU, assuming that the disk
radiates like a blackbody. 2a) Each curve in this figure describes the
temperature dependence vs radius calculated using the dissipation energy
mechanism indicates (eq. [\ref{Tcalc}]).The values attributed to $T_{irr}$, the
temperature for an irradiated disk model was extracted from Figure 3
of D'Alessio et al. (1998); 2b) In this figure, we show the total 
temperatures reached in our models, taking into account only the viscous
dissipation. They are obtained by summing the energy radiated by
corresponding to viscous dissipation with the Alfvenic heatings 
(equations 29-32). 
2c) This figure shows the total temperatures summing viscous dissipation, 
the irradiated model and each Alfvenic heating. \label{fig2}}

\figcaption[f3a,b.eps] {Temperatures associated with nonlinear Alfvenic
dissipation, calculated for: 3a) for region 3 (1.5 AU $\leq$ R $\leq$ 5 AU) and
3b) for region 4 (5 AU $\leq$ R $\sim$ 100 AU). \label{fig3}}

\figcaption[f4a,b,c.eps]{Dissipated energies (in units of erg ${\rm
cm}^{-2}$ ${\rm s}^{-1}$) as a function of disk radius (in AU),
obtained from the turbulent heating model, for: 4a) region 2 (0.1 $\leq$ 
R $\leq$ 1.4 AU), 4b) region 3 (1.5 AU $\leq$ R $\leq$ 5 AU) and 4c) region 4
(5 AU $\leq$ R $\sim$ 100 AU).
We observe that turbulent Alfvenic heating is much greater
than viscous heating for all density profiles.  \label{fig4}}


\begin{deluxetable}{cccccc}
\footnotesize
\tablecaption{Parameters of a stationary, geometrically thin, 
optically thick protostellar
accretion disk, with the opacity law of Bell \& Lin (1994) in the 
iron-silicate regime. \label{tab1}}
\tablehead{
\colhead{R (AU)} & \colhead{$T_{c}$ (K)} & \colhead{$T_{d}$ (K)} & 
\colhead{h (AU)} & \colhead{$\Sigma$ (g ${\rm cm}^{-2})$} & 
\colhead{$\rho_{c}$ (g ${\rm cm}^{-3})$}} 
\startdata
$0.1$ & $1362.36$ & 369.28 & 0.008 & 66.92 & $5.31 \times 10^{-10}$ \nl
$0.2$ & 718.80 & 226.31 & 0.017 & 50.61 & $1.95 \times 10^{-10}$ \nl
0.3 & 489.80 & 169.04 & 0.026 & 42.47 & $1.08 \times 10^{-10}$ \nl
$0.4$ & 371.99 & 137.20 & 0.035 & 37.36 & $7.09 \times 10^{-9}$ \nl
0.5 & 300.09 & 116.60 & 0.044 & 33.77 & $5.11 \times 10^{-9}$ \nl
0.6 & 251.60 & 102.05 & 0.053 & 31.06 & $3.90 \times 10^{-9}$ \nl
0.7 & 216.67 & 91.15 & 0.062 & 28.92 & $3.11 \times 10^{-9}$ \nl
0.8 & 190.29 & 82.63 & 0.071 & 27.18 & $2.55 \times 10^{-9}$ \nl
0.9 & 169.67 & 75.78 & 0.080 & 25.73 & $2.14 \times 10^{-9}$ \nl
1.0 & 153.09 & 70.12 & 0.089 & 24.48 & $1.83 \times 10^{-9}$ \nl
1.1 & 139.48 & 65.36 & 0.098 & 23.41 & $1.59 \times 10^{-9}$ \nl
1.2 & 128.10 & 61.30 & 0.107 & 22.47 & $1.40 \times 10^{-9}$ \nl
1.3 & 118.45 & 57.78 & 0.116 & 21.63 & $1.24 \times 10^{-9}$ \nl
1.4 & 110.15 & 54.71 & 0.126 & 20.88 & $1.11 \times 10^{-9}$ \nl
\enddata
\tablecomments{Here, $R$ is the disk radius, $T_{c}$ is the temperature
of the midplane of the disk, $T_{d}$ is the temperature due to viscous
dissipation, $h$ is the disk semi-thickness, $B$ is the magnetic field
(obtained from the model of Gammie 1996) and $\rho_{c}$, the
midplane density.} 
\end{deluxetable}

\begin{deluxetable}{cccccc}
\footnotesize
\tablecaption{Parameters of the disk calculated using the layered 
model of Gammie (1996), for a region extending from 0.1 AU to
1.4 AU. \label{tab2}}
\tablehead{
\colhead{R (AU)} & \colhead{$T_{c}$ (K)} & \colhead{$T_{d}$ (K)} & 
\colhead{h (AU)} & \colhead{B (G)} & 
\colhead{$\rho_{c}$ (g ${\rm cm}^{-3})$}}
\startdata
$0.1$ & $1005.06$ & $302.64$ & $0.003$ & $4.96$ & $3.51 \times 10^{-9}$ \nl
$0.2$ & $663.09$ & $210.32$ & $0.008$ & $2.66$ & $1.53 \times 10^{-9}$ \nl
0.3 & 519.90 & 169.99 & 0.012 & 1.84 & $9.40 \times 10^{-10}$ \nl
$0.4$ & $437.48$ & $146.16$ & $0.017$ & $1.42$ & $6.66 \times 10^{-10}$ \nl
0.5 & 382.66 & 130.01 & 0.023 & 1.16 & $5.09 \times 10^{-10}$ \nl
$0.6$ & $343.01$ & $118.14$ & $0.028$ & $0.99$ & $4.09 \times 10^{-10}$ \nl
0.7 & 312.70 & 108.95 & 0.034 & 0.86 & $3.40 \times 10^{-10}$ \nl
$0.8$ & $288.63$ & $101.58$ & $0.040$ & $0.76$ & $2.90 \times 10^{-10}$ \nl
0.9 & 268.93 & 95.49 & 0.046 & 0.69 & $2.51 \times 10^{-10}$ \nl
$1.0$ & $252.46$ & $90.35$ & $0.052$ & $0.62$ & $2.22 \times 10^{-10}$ \nl
1.1 & 238.43 & 85.94 & 0.059 & 0.57 & $1.98 \times 10^{-10}$ \nl
1.2 & 226.30 & 82.10 & 0.065 & 0.53 & $1.78 \times 10^{-10}$ \nl
1.3 & 215.69 & 78.72 & 0.072 & 0.49 & $1.62 \times 10^{-10}$ \nl
$1.4$ & $206.31$ & $75.72$  & $0.078$ & $0.46$ & $1.48 \times 10 ^{-10}$ \nl
\enddata
\tablecomments{In this table, $R$ is the disk radius, $T_{c}$ is the
temperature of the midplane of the disk, $T_{d}$ is the temperature due
to viscosity dissipation, $h$ is the semi-thickness of the disk, $B$ is
the magnetic field and $\rho_{c}$, the midplane density.}
\end{deluxetable}

\end{document}